\documentclass[conference]{IEEEtran}
\IEEEoverridecommandlockouts
\usepackage{cite}
\usepackage{amsmath,amssymb,amsfonts}
\usepackage{algorithmic}
\usepackage{graphicx}
\usepackage{textcomp}
\usepackage{xcolor}
\usepackage{multirow}
\usepackage{url}
\def\BibTeX{{\rm B\kern-.05em{\sc i\kern-.025em b}\kern-.08em
    T\kern-.1667em\lower.7ex\hbox{E}\kern-.125emX}}
\begin{document}

\title{Structural Watermarking to Deep Neural Networks via Network Channel Pruning}

\author{Xiangyu Zhao, Yinzhe Yao, Hanzhou Wu$^*$ and Xinpeng Zhang\\
Shanghai University, Shanghai 200444, P. R. China\\
$^*$h.wu.phd@ieee.org
}

\maketitle

\begin{abstract}
In order to protect the intellectual property (IP) of deep neural networks (DNNs), many existing DNN watermarking techniques either embed watermarks directly into the DNN parameters or insert backdoor watermarks by fine-tuning the DNN parameters, which, however, cannot resist against various attack methods that remove watermarks by altering DNN parameters. In this paper, we bypass such attacks by introducing a structural watermarking scheme that utilizes channel pruning to embed the watermark into the host DNN architecture instead of crafting the DNN parameters. To be specific, during watermark embedding, we prune the internal channels of the host DNN with the channel pruning rates controlled by the watermark. During watermark extraction, the watermark is retrieved by identifying the channel pruning rates from the architecture of the target DNN model. Due to the superiority of pruning mechanism, the performance of the DNN model on its original task is reserved during watermark embedding. Experimental results have shown that, the proposed work enables the embedded watermark to be reliably recovered and provides a sufficient payload, without sacrificing the usability of the DNN model. It is also demonstrated that the proposed work is robust against common transforms and attacks designed for conventional watermarking approaches.
\end{abstract}

\begin{IEEEkeywords}
Watermarking, deep neural networks, ownership protection, deep learning, security.
\end{IEEEkeywords}

\section{Introduction}
\IEEEPARstart{D}{ue} to the superior performance of deep neural networks (DNNs), they have become the default choice for solving difficult problems in various areas such as computer vision \cite{VGGNet} and natural language processing \cite{BERT}. However, in practice, the creation of a high-performance DNN model is very expensive, which requires huge amounts of well labelled data, expertise of designing and training the model, and substantial computational resources. Therefore, because of the commercial value and the high cost of building DNN models, it is very common to consider DNN models as valuable products, highlighting the importance of protecting DNN models from intellectual property (IP) infringement.

As an effective means to IP protection, digital watermarking enables the owner to inject a unique signature containing the copyright information into the host signal secretly. Inspired by watermarking systems designed for classical media signals such as images and videos, recently, increasing watermarking techniques are proposed to protect DNN models. These works can be roughly divided to two categories, i.e., \emph{white-box watermarking} and \emph{black-box watermarking}. White-box approaches treat DNNs as static signals and embed watermarks into the noise-like components of the host signals, e.g., \cite{Uchida2017Paper, wang:idnn, Passports, Chen:deepmarks, ICPR:2021}. Black-box approaches take advantage of the functionality of DNNs and use specifically crafted samples to fine-tune the host DNN, e.g., \cite{deepsigns, adi:paper, adv:paper, zhao:wgnn, wu:wdnn}. As a result, the watermarks are embedded into the unique predictions of the host DNN to the special samples. In general, most of existing works have one thing in common that they alter the DNN parameters for embedding watermarks.

On the other hand, the vulnerability of parameter-based watermarking techniques have been extensively studied. Firstly, some works \cite{wnnAttack1, wnnAttack2, wnnAttack3} show that similar to conventional media watermarking, embedding watermarks into DNN weights may introduce abnormal properties to statistical characteristics of the weights, making watermarks detectable and hence removable. Secondly, an adversary can remove black-box watermarks by carefully fine-tuning the DNN parameters (weights) \cite{remove1, remove2, remove3, remove4, remove5}. Generally speaking, most of the existing attacks are performed by intentionally changing DNN parameters for removing watermarks embedded by crafting the parameters.

Based on the aforementioned insights, a reasonable solution for robust watermarking should be embedding a watermark in the network structure, rather than the network parameters, to resist against the above attacks. Recently, Lou \emph{et al.} \cite{NAS:wdnn} propose a method to embed a watermark in the model architecture leveraging neural architecture search (NAS) \cite{NAS}. In the work, certain connections with specific operations are fixed in the search space controlled by the watermark. The remaining connections are then searched to improve network performance. In other words, the secret watermark is embedded during the model building phase. In practice (e.g., watermarking DNN can be outsourced), it is sometimes desirable to watermark an existing trained model, rather than watermarking in the process of creating a model from scratch, which often requires a lot of computational resources.

In this paper, we propose a novel method to exploit network pruning for \emph{structural embedding}, which allows us to insert a watermark to a given DNN. Network pruning is a very popular approach to reduce the DNN size by pruning the redundant components, which plays a quite important role in constructing lightweight DNNs. Inspired by this technique, we embed a watermark into the redundant architecture of the given DNN, ensuring that the performance of the DNN on its original task can be kept well. Specifically, we prune the internal channels of the host DNN with the channel pruning rates controlled by the watermark. During watermark extraction, the watermark is retrieved by identifying the channel pruning rates from the architecture of the target DNN model. Our experiments show that, thanks to network pruning, the proposed method enables the embedded watermark to be reliably recovered and provides a sufficient payload, without sacrificing the usability of the DNN model. It is also demonstrated that the proposed method is robust against common transforms and attacks designed for conventional watermarking methods. 

The main contributions of this paper are as follows:

\begin{itemize}
	\item To the best knowledge of the authors, it is the first work incorporating network pruning into DNN watermarking.
	\item The proposed method embeds a secret watermark into the model architecture, rather than parameters, and can thus resist most existing transforms and attacks against DNN watermarking.
	\item Extensive experiments indicate that the proposed method achieves superior performance in IP protection while does not impair the original task of the DNN.
\end{itemize}

The rest of this paper will be organized as follows. We first introduce preliminaries of channel pruning in Section II. Then, in Section III, we introduce the proposed DNN watermarking method. We provide experiments in Section IV to evaluate the proposed method, followed by conclusion in Section V. 

\section{Network Channel Pruning}
DNN models usually contain huge quantities of parameters and complex structures, which limit their value in practical applications. As an important branch of network compression, many network pruning techniques have been studied to reduce the size of DNN models while maintaining the performance on the primary task, which bases on the observation that redundant parameters or architectures actually exist in DNNs. By exploiting and removing relatively unimportant components in DNNs, the storage space and running memory are reduced and the calculating speed is accelerated, which greatly improves the usability of DNNs in practical scenarios, e.g., deploying such lightweight DNN models in the mobile terminals. 

Network pruning \cite{Pruning:Survey} can be roughly divided to three categories: weight-level pruning, channel-level pruning and layer-level pruning. The weight-level pruning sets certain weights to zero and does not change the structure. Since our purpose is to craft network structure for watermarking, weight-level pruning is excluded from our candidates first. Compared with layer-level pruning, we select channel pruning as the final strategy because pruning channels is more concealed and the number of channels in a network is much more than that of layers, which reserves more space for embedding extra information. 

A channel pruning technique typically consists of the following steps: 1) evaluate the importance of the channels, 2) sort the channels according to the importance and then prune $\alpha\%$ (pruning rate) least important channels, 3) fine-tune the compressed model using the training dataset to recover the performance. Note that literally all the channels in a network are sorted together for pruning. While in the proposed method, we perform channel pruning layer by layer locally rather than pruning globally. In other words, we set different pruning rates for different layers and only the channels inside the same layer are compared together and then pruned given the pruning rate of the layer.

In this paper, we incorporate two novel channel pruning schemes into our watermark solution due to their simplicity and efficiency: 1) \emph{Network Slimming} proposed in \cite{ZhuangLiu:Pruning}, which prunes channels with small scaling factors in the batch normalization (BN) \cite{bn} layers. 2) \emph{L1-norm Pruning} \cite{HaoLi:Pruning} prunes the channels based on the L1-norm of weights of convolutional layers. We show more details of the two pruning techniques below. It is pointed that, the proposed framework is general and does not rely on any particular pruning technique. 

\subsection{Network Slimming}
Network slimming \cite{ZhuangLiu:Pruning} directly utilizes the scaling factors of the following/previous BN layer as the importance of the output/input channels of a convolutional layer. The channels are then pruned according to the scaling factors. To be specific, BN normalizes and transforms the scale of the internal activations in a batch to improve training efficiency. Let $z_\text{in}$ and $z_\text{out}$ be the input and output activations, a BN layer can be formulated as:
\begin{equation}
\hat{z} = \frac{z_\text{in} - \mu_\text{B}}{\sqrt{{\sigma^2_\text{B}} + \epsilon}}\quad\text{and}\quad z_\text{out} = \gamma\cdot\hat{z} + \beta,
\end{equation}
where $\mu_\text{B}$ and $\sigma^2_\text{B}$ represent the mean and variance of the input activations over the batch $\text{B}$, and $\epsilon$ is a very small positive real number. So, the first equation normalizes the input activations. In the second equation, $\gamma$ and $\beta$ are trainable parameters that linearly transform the normalized activations $\hat{z}$ into a new scale that is better suited for learning. It is common in practice that each convolutional layer is followed by a BN layer and the distribution transform is performed in a channel-wise manner, meaning that each channel of the convolutional layer is multiplied by a scaling factor $\gamma$ in the BN layer. So a relatively small $\gamma$ indicates that the corresponding channel is unimportant after the multiplication. Then, $\alpha \%$ (pruning rate) of channels with smallest scaling factors can be pruned. 

In the original version of network slimming, a sparsity-induced penalty term is added to the training objective to achieve sparsity of scaling factors for better distinguishing importance of channels. We ignore this strategy since we aim to mark an existing well-established model. It indicates that the model pruning procedure should be independent of the model training procedure. 

\subsection{L1-norm Pruning}
Unlike network slimming that requires the weights of BN layers, L1-norm pruning \cite{HaoLi:Pruning} prunes the channels of convolutional layers based on the weights of convolutional layers independent of other layers, which is more general because some convolutional layers may not be followed by BN layers. 

In detail, the parameters of a convolutional layer are organized into a tensor $\textbf{F}$ with size $c_\text{out} \times c_\text{in} \times w \times h$, where $c_\text{in}$ and $c_\text{out}$ are the input channel size and the output channel size, respectively, $w \times h$ is the shape of kernels. The parameter tensor can be expanded along the dimension of the output channel to obtain $c_\text{out}$ filters with size $c_\text{in} \times w \times h$ so each filter actually corresponds to an output channel. Then, the relative importance of a channel is measured by the L1-norm of the corresponding filter based on the assumption that small weights produce unimportant channels. The L1-norm of the $k$-th filter is expressed as $|| \textbf{F}_{k,*,*,*} ||_1 = \sum_{a,b,c} |\textbf{F}_{k,a,b,c}|$, which actually represents the expected magnitude of produced feature maps of the related channels. After computing the L1-norm of all filters in the convolutional layer, the channels are sorted and pruned. Finally, the pruned network is fine-tuned.

\section{Proposed Method}
The proposed watermarking framework has two phases: \emph{watermark embedding} and \emph{watermark extraction}. The purpose of watermark embedding is to insert a sequence of bits containing ownership information into the host model. In this work, we embed watermark bits into the pruning rates during model pruning by leveraging quantization index modulation (QIM) \cite{QIM}. During embedding, we partition the watermark into several bit-segments and use each bit-segment to decide a pruning rate, where the pruning rate is actually quantized with the bit-segment. Then each generated pruning rate is assigned to a convolutional layer for channel pruning controlled by the secret key. Once the host model is marked, it will be put into use. The performance of the marked model on its original task will not be degraded due to the efficiency of pruning. During ownership verification, the watermark can be reliably reconstructed by determining the channel pruning rates.

\subsection{Watermark Embedding}
Let $W = \{w_1, w_2, ..., w_n\} \in \{0, 1\}^n$ denote the watermark to be embedded. We partition $W$ into multiple bit-segments. In other words, we can rewrite $W$ as $W = \{W_1, W_2, ..., W_m\}$, where $m$ is the number of bit-segments and for any $1\leq i\leq m$, we have $W_i = \{w_{(i-1)l+1}, w_{(i-1)l+2}, ..., w_{il}\}$ and $n = ml$. In case that $l$ does not divide $n$, we can always append ``0'' to the end of the watermark in advance such that $l$ divides $n$. Therefore, we consider $n=ml$ by default. On the other hand, all convolutional layers of the host DNN are collected to constitute a convolutional-layer set, which is expressed as $C$ = $\{c_1, c_2, ..., c_t\}$, where $t$ should be no less than $m$ so that $W$ can be fully carried by pruning $C$ because each $c_i\in C$ will be used to carry at most one bit-segment of $W$. Obviously, according to a secret key, we can construct a subset of $C$, i.e., $C_\text{cov} = \{c_{s_1}, c_{s_2}, ..., c_{s_m}\}$, where $1\leq s_1, s_2, ..., s_m \leq t$. Each bit-segment $W_i$ will be carried by the pruning rate applied to $c_{s_i}$. It means that the (channel) pruning rate for $c_{s_i}$, denoted by $p_{s_i}\in [0, 1)$, can be determined according to $W_i$. By applying $\{p_{s_1}, p_{s_2}, ..., p_{s_m}\}$ to $C_\text{cov}$ during pruning, we can get the (marked) pruned subset $C_\text{ste} = \{c_{s_1}', c_{s_2}', ..., c_{s_m}'\}$. 
It is free to prune $C\setminus C_\text{cov}$ as $C\setminus C_\text{cov}$ does not carry information. 

We limit the pruning rate for each convolutional layer to be pruned to the range $[p_\text{min}, p_\text{max})$, where $0 \leq p_{min} < p_{max} \leq 1$. We do not use the default range $[0, 1)$ because we hope to limit the reduction of the DNN performance on its original task due to pruning by limiting the pruning range. We partition the range $[p_\text{min}, p_\text{max})$ into $2^l$ sub-ranges, namely, $[p_\text{min}, p_\text{max}) = \cup_{i=0}^{2^l-1} [p_\text{min}+i\cdot\Delta, p_\text{min}+(i+1)\cdot\Delta)$, where $\Delta = (p_\text{max} - p_\text{min}) / 2^l$. Given $W_i$, the corresponding pruning rate $p_{s_i}$ will be determined as $p_{s_i} = p_\text{min} + (d_i + 0.5)\cdot\Delta$,  where $d_i$ represents the decimal value of $W_i$, i.e., $d_i = \sum_{j=1}^{l}w_{(i-1)l + j}\cdot 2^{l-j}$. In this way, with the pruning rates carrying the watermark, the marked DNN model can be produced by channel pruning. 

\subsection{Watermark Extraction}
The watermark extraction procedure is to determine whether a watermark exists in a target DNN model. If the watermark can be precisely extracted, the target model is demonstrated to be a pirate version of the host DNN. Because the mapping between each bit-segment of the watermark and the pruning rate of the corresponding channel is actually bijective, the watermark can be recovered by checking the pruning rates of the marked channels. For example, given $p_{s_i}$, we first determine the decimal value of $W_i$ as $d_i = \lfloor (p_{s_i} - p_\text{min}) / \Delta \rfloor$, where $\lfloor x \rfloor$ represents the largest integer no more than $x$. Then, the binary version of $W_i$ can be easily determined from $d_i$. By collecting all the binary sequences for $\{W_1, W_2, ..., W_m\}$, the entire watermark $W$ can be reconstructed, which will be used to identify ownership. It is pointed that, one should compare the target (marked) DNN model and the original non-marked DNN model to determine the channel pruning rates. Therefore, the proposed work is a \emph{non-oblivious} watermarking system. 

The proposed framework can be easily extended to the blind watermarking design. The core idea of the proposed work is to mark the network structure by extracting statistics as a 1-D cover sequence for watermarking. On this basis, any blind watermarking operation can be applied to a well-built cover sequence, resulting in ``\emph{blind structural DNN watermarking}''.

\section{Experimental Results and Analysis}
In this section, we show experimental results and analysis to verify the superiority of the proposed work. Our simulation bases on two open source model pruning projects\footnote{\url{https://github.com/foolwood/pytorch-slimming}}$^,$\footnote{\url{https://github.com/Eric-mingjie/rethinking-network-pruning}}.

\subsection{Setup}
We conduct extensive experiments on the CIFAR-10 dataset and the CIFAR-100
dataset\footnote{\url{https://www.cs.toronto.edu/~kriz/cifar.html}}. The two datasets are composed of natural images with resolution $32\times32$, where the CIFAR-10 contains 10 classes while the CIFAR-100 contains 100 classes. Both consist of 60,000 samples, among which the training set and testing set have 50,000 and 10,000 samples, respectively.

Since we here mainly focus on watermarking convolutional neural networks (CNNs) via channel pruning, we use three classical architectures as our basis: VGGNet \cite{VGGNet}, ResNet \cite{ResNet} and DenseNet \cite{DenseNet}. For VGGNet, we use the popular variation with 19 layers. We also use 164-layer ResNet with bottleneck residual blocks. As for DenseNet, the architecture consists of 40 layers and the growth rate is set to be 12. We term the three DNN architectures as VGG-19, ResNet-164 and DenseNet-40. 
Before evaluating the proposed watermarking system, we train the host networks normally. For both datasets, we train 200 epochs for VGG-19 and 160 epochs for ResNet-164 and DenseNet-40. During training, we optimize all the networks by using stochastic gradient descent (SGD) with learning rate 0.01. We also use a weight decay of $10^{-4}$ and augment the training data by mirroring and shifting to avoid overfitting. In summary, our training settings closely follow \cite{ZhuangLiu:Pruning}. 

As mentioned in the previous section, we use two effective channel pruning techniques as our basis for watermarking. For ResNet-164 and DenseNet-40, due to the residual connections, the output feature map of a layer can be fed to multiple layers and the BN layers are set before the corresponding convolutional layers. So the channels are actually pruned at the front end of layers. In order to align the shapes of tensors in the network, we set channel selection layers that mask out unimportant channels after the BN layers that receive multiple incoming activations from residual connection. Such strategy was first proposed in \cite{ZhuangLiu:Pruning}. We ignore the convolution kernels in the residual connections during pruning for convenience. In all our watermarking procedures, we limit all pruning rates to the range $[0, 0.7)$, which not only provides much space for watermarking but also confines the degree of pruning so that the performance can be well recovered.

After watermarking by pruning, we have to fine-tune the marked DNN to improve the performance on the original task. We fine-tune both ResNet-164 and DenseNet-40 with 20 epochs (using the original training set with a learning rate of 0.001). For VGGNet-19, we fine-tune it with 40 epochs since it requires more epochs of training till convergence. Other fine-tuning settings is the same as the original model training.

\begin{table}[!t]
	\centering
	\caption{Watermarking channel rates of different network architectures due to various parametric settings.}
	\label{TableI}
	\renewcommand{\arraystretch}{1.25}
	\scalebox{0.72}{
		\begin{tabular}{|c|c|c|c|c|c|c|c|}
			\hline
			\multirow{2}{*}{\begin{tabular}{c}
					DNN
			\end{tabular}} & \multirow{2}{*}{ $t$ }	& \multirow{2}{*}{ $l$ }	
			& \multicolumn{5}{c|}{ Watermarking Channel Rate (bits) } \\ \cline{4-8}
			& & & $r_\text{cov} = 0.2$ & $r_\text{cov} = 0.4$ & $r_\text{cov} = 0.6$ & $r_\text{cov} = 0.8$ & $r_\text{cov} = 1.0$ \\ \hline
			\multirow{3}{*}{VGG-19} & \multirow{3}{*}{16}
			& $1$ & 3 & 6 & 9 & 12 & 16 \\ \cline{3-8}
			& & $2$ & 6 & 12 & 20 & 26 & 32 \\ \cline{3-8}
			& & $3$ & 9 & 18 & 30 & 39 & 48 \\ \cline{1-8}
			
			\multirow{3}{*}{ResNet-164} & \multirow{3}{*}{162}
			& $1$ & 32 & 65 & 97 & 130 & 162 \\ \cline{3-8}
			& & $2$ & 64 & 130 & 194 & 260 & 324 \\ \cline{3-8}
			& & $3$ & 96 & 195 & 291 & 390 & 486 \\ \cline{1-8}
			
			\multirow{3}{*}{DenseNet-40} & \multirow{3}{*}{39}
			& $1$ & 8 & 16 & 23 & 31 & 39 \\ \cline{3-8}
			& & $2$ & 16 & 32 & 46 & 62 & 78 \\ \cline{3-8}
			& & $3$ & 24 & 48 & 69 & 93 & 117 \\ \cline{1-8}
			
		\end{tabular}
	}
\end{table}

\begin{table}[!t]
	\centering
	\caption{Classification precision due to different $l$. Here, $r_\text{cov} = 1.0$.}
	\label{TableII}
	\renewcommand{\arraystretch}{1.25}
	\scalebox{0.75}{
		\begin{tabular}{|c|c|c|c|c|c|c|}
			\hline
			\multirow{2}{*}{Dataset} & \multirow{2}{*}{
				\begin{tabular}{c}
					DNN
				\end{tabular}
			}	
			& \multirow{2}{*}{Baseline}
			& \multirow{2}{*}{\begin{tabular}{c}
					Pruning \\ Scheme
			\end{tabular}}
			& \multicolumn{3}{c|}{Top-1 Precision (marked)} \\ \cline{5-7}
			& & & & $l = 1$
			& $l = 2$
			& $l = 3$ \\ \cline{5-7}
			\hline
			\multirow{6}{*}{CIFAR-10} 
				& \multirow{2}{*}{VGG-19} & \multirow{2}{*}{92.70\%} 
				& Network Slimming & 92.04\% & 92.64\% & 91.96\% \\ \cline{4-7}
				& & & L1-norm Pruning & 92.19\% & 92.72\% & 91.90\% \\ \cline{2-7}
				
				& \multirow{2}{*}{ResNet-164} & \multirow{2}{*}{94.92\%} 
				& Network Slimming & 93.25\% & 93.17\% & 93.11\% \\ \cline{4-7}
				& & & L1-norm Pruning & 94.02\% & 93.94\% & 94.16\% \\ \cline{2-7}
				
				& \multirow{2}{*}{DenseNet-40} & \multirow{2}{*}{93.90\%} 
				& Network Slimming & 92.47\% & 91.10\% & 92.52\% \\ \cline{4-7}
				& & & L1-norm Pruning & 93.96\% & 93.63\% & 93.80\% \\ \cline{2-7}
				\hline
			
			\multirow{6}{*}{CIFAR-100} 
				& \multirow{2}{*}{VGG-19} & \multirow{2}{*}{70.00\%} 
				& Network Slimming & 68.73\% & 68.83\% & 68.11\% \\ \cline{4-7}
				& & & L1-norm Pruning & 68.97\% & 68.59\% & 69.62\% \\ \cline{2-7}
				
				& \multirow{2}{*}{ResNet-164} & \multirow{2}{*}{76.80\%} 
				& Network Slimming & 72.63\% & 71.95\% & 72.21\% \\ \cline{4-7}
				& & & L1-norm Pruning & 74.77\% & 74.10\% & 74.03\% \\ \cline{2-7}
				
				& \multirow{2}{*}{DenseNet-40} & \multirow{2}{*}{74.59\%} 
				& Network Slimming & 68.59\% & 67.18\% & 68.87\% \\ \cline{4-7}
				& & & L1-norm Pruning & 72.05\% & 71.34\% & 72.58\% \\ \cline{2-7}
				\hline
		\end{tabular}
	}
\end{table}

\subsection{Watermarking Channel Rate}
In this subsection, we look into the volume of watermark information that can be embedded under our watermarking framework. The maximum number of embeddable bits (i.e., watermarking channel rate) is $n_\text{max} = lt$. So, a relatively deeper DNN is potentially more suited for embedding more information. In practice, embedding in certain randomly selected layers rather than all the layers gives more stealthiness for the watermark. To normalize across different network architectures, we term the proportion of pruned layers as $r_\text{cov} = |C_\text{cov}| / t$, where $|C_\text{cov}|$ is the number of layers with channels pruned. Given $r_\text{cov}$, the watermarking channel rate is $N = ltr_\text{cov}$ (bits). Table I shows the channel rates of different networks under different $r_\text{cov}$ and $l$. It can be inferred that, to embed more secret bits, we should use larger $l$ and/or $r_\text{cov}$ for a given host network. However, a larger $l$ means that the range of pruning rate is segmented into more parts so that the length of each part is smaller, making the embedded watermark less robust against perturbations. On the other hand, a larger $r_\text{cov}$ could reduce the stealthiness of the watermark and improve the difficulty of recovering the network performance on its original task. Therefore, in practice, it is necessary for us to make a trade-off between channel rate and embedding parameters. To this end, we show more experiments and analysis below.

\subsection{Watermark Reliability}
Watermark reliability requires the embedded watermark to be correctly extracted from the marked model for ownership verification. It is obvious that our watermarking scheme is essentially reliable because the mapping between each bit-segment of the watermark and the pruning rate of the corresponding channel is actually bijective. As a result, as long as the architecture of the watermarked model is not modified, the watermark can be precisely extracted given the secret key.

\begin{figure*}[!t]
	\centering
	\includegraphics[width=\linewidth]{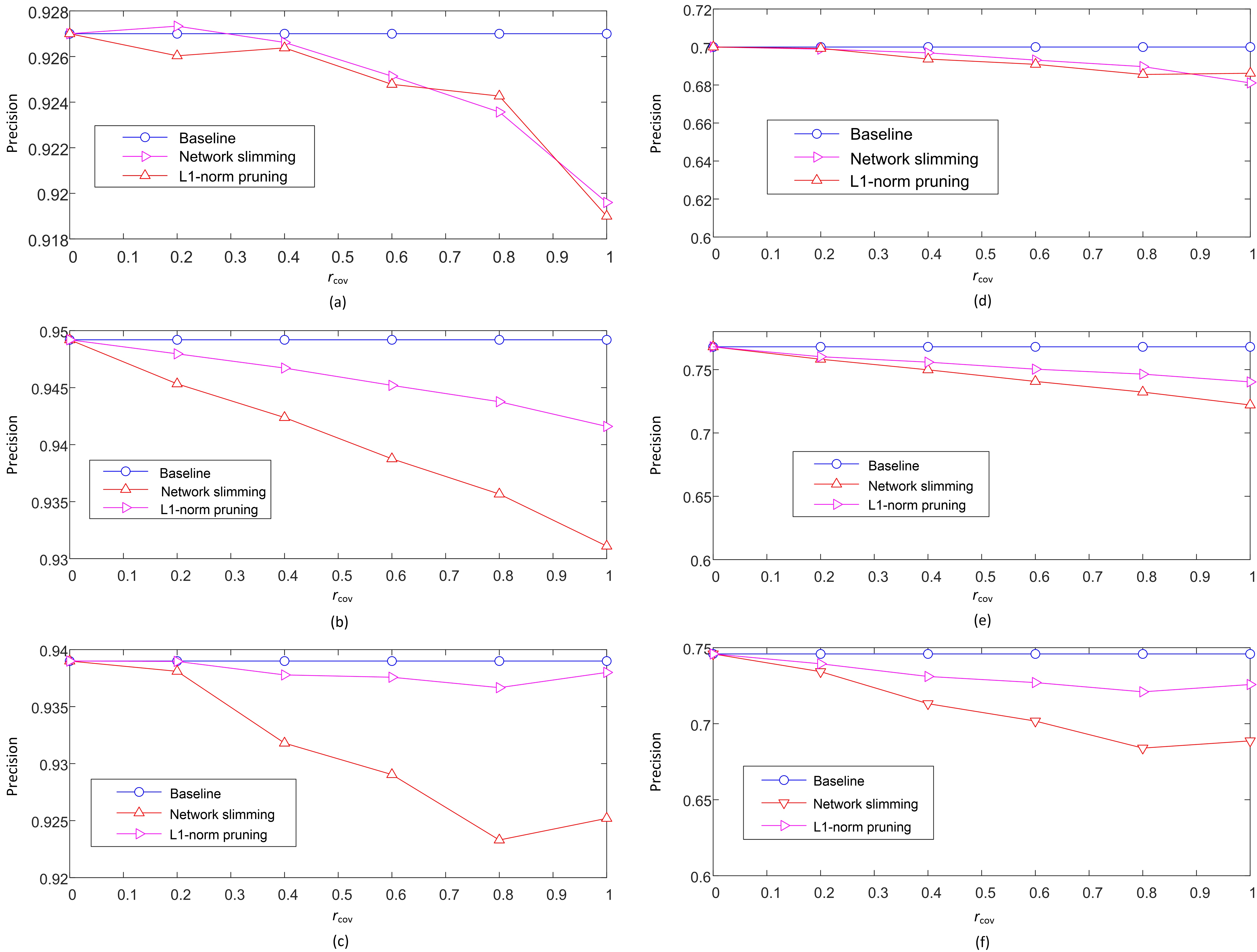}\\
	\caption{Precision due to different $r_\text{cov}$. The precision in the figure is averaged across precisions of watermarked models produced by different $l$. (a) CIFAR-10, VGG-19, (b) CIFAR-10, ResNet-164, (c) CIFAR-10, DenseNet-40, (d) CIFAR-100, VGG-19, (e) CIFAR-100, ResNet-164, (f) CIFAR-100, DenseNet-40.}
\end{figure*}

\subsection{Network Fidelity}
Fidelity means that the performance of the marked (pruned) network should not be significantly degraded. Our watermarking procedure is actually a pruning procedure so that the degree of performance degradation is intuitively proportional to the size of the pruned component, which is controlled by the pruning rates $\{p_{s_1}, p_{s_2}, ..., p_{s_m}\}$. Let $A$ = $\{a_0, a_1, ..., a_{2^l-1}\}$, where $a_i = p_\text{min}+(i+0.5)\cdot\Delta$ for all $0\leq i\leq 2^l-1$, denote the set including all possible values of a pruning rate. It can be determined that the expectation of a pruning rate across all possible states of a bit-segment is $\mathbb{E}_p = \sum_{i=0}^{2^l-1}\frac{1}{2^l}a_i = \frac{1}{2}(p_\text{min}+p_\text{max})$, which indicates that the expectation of the pruning rate equals the middle point of $[p_\text{min}, p_\text{max})$ and is independent of $l$. Therefore, given the pruning range $[p_\text{min}, p_\text{max})$, the entire pruning rate of the host network during watermarking is mainly affected by $r_\text{cov}$. It other words, the network fidelity mainly relies on $r_\text{cov}$. To demonstrate our analysis, we train the host non-marked models and their marked versions. We evaluate their performance on the testing datasets.  

We first evaluate the impact of $r_\text{cov}$ on fidelity. It can be seen in Fig. 1 that for both datasets and all models, the performance of watermarked model will decline as $r_\text{cov}$ increases. We can also observe that for networks with residual connections, i.e., ResNet-164 and DenseNet-40, the performance recovery of networks watermarked by L1-norm pruning is better than those watermarked by BN pruning (i.e., network slimming). We believe the reason is that we ignore the sparsity penalty term that sparsifies BN scaling factors so that the channel importance produced by unsparsified BN weights is less accurate than that evaluated by L1-norm of filters. But note that L1-norm pruning is comparatively more expensive, which requires the computation of L1-norm while BN pruning only needs to look into the BN weights. 

We also investigate the effect of $l$. Specifically, we set $r_\text{cov}$ to be 1.0 (every layers are watermarked) and test the watermarked models due to different $l$. The results are shown in Table II. It can be seen that even under the worst case ($r_\text{cov}=1.0$) although the performance is degraded after watermarking but the degradation can be kept in a very low level. This is due to the effectiveness of pruning techniques. We can also see that generally the precision is independent of $l$, which matches our analysis.

The above experiments have demonstrated that, in terms of network fidelity, the network performance (on the original task) can be well recovered under our watermarking framework under proper settings. In order to embed more information without sacrificing the network fidelity, we can set a relatively large $l$ and a relatively small $r_\text{cov}$, which means to embed more bits into less layers. However, it should be admitted that, as illustrated before, embedding more bits into a layer reduces the watermark robustness against structural modifications.

\subsection{Robustness}
Robustness is an important criterion for evaluating a watermarking solution. In practical applications, the watermarked content may be normally transformed for usage or even deliberately attacked for watermark removal. Robustness is the probability of recovering the watermark after normal transform and watermark removal attack.

Model parameter pruning and fine-tuning are two most considered modifications against deep learning models. It is clear that our watermarking solution can perfectly resist against parameter pruning and fine-tuning because we mark the network in the structure so that these parameter-level modifications cannot impair the watermark at all. Despite that, as illustrated before, most attacks for removing watermarks are also parameter-based, which cannot break through our watermarking. Indeed, some attacks like knowledge distillation \cite{arXiv:Distilling} change the model structure and are thus capable of removing our watermarks. But such mechanisms often require too much data and computational resources. Given the resources, training a new model from scratch would be a better choice for the adversary. 

\section{Conclusion and Discussion}
In this paper, we present a novel structural-based network watermarking technique by incorporating channel pruning. In the work, we use the watermark to determine the pruning rates of channels for pruning. After pruning and fine-tuning the network, the watermark is embedded in the network architecture, which can be identified by checking the pruning rates during verification. Experimental results and theoretical analysis show the superiority of the proposed method in terms of network fidelity, watermarking channel rate, watermark reliability and robustness and security. We believe watermarking in network structure shows particular advantages compared with traditional watermarking methods, which has great potential for intellectual property protection of deep neural networks. In the future, we will explore more ways of watermarking into different levels of architectures.  

\section*{Acknowledgement}
This work was supported in part by the National Natural Science Foundation of China (NSFC) under Grant 61902235 and Grant U1936214; and in part by the Shanghai ``Chen Guang'' Program under Grant 19CG46; and in part by the Science and Technology Commission of Shanghai Municipality (STCSM) under Grant 21010500200. 


\end{document}